\documentclass{article}

\usepackage{arxiv}

\usepackage[utf8]{inputenc} % allow utf-8 input
\usepackage[T1]{fontenc}    % use 8-bit T1 fonts
\usepackage{hyperref}       % hyperlinks
\usepackage{url}            % simple URL typesetting
\usepackage{booktabs}       % professional-quality tables
\usepackage{amsfonts}       % blackboard math symbols
\usepackage{nicefrac}       % compact symbols for 1/2, etc.
\usepackage{microtype}      % microtypography
\usepackage{lipsum}		% Can be removed after putting your text content
\usepackage{graphicx}
\usepackage{natbib}
\usepackage{doi}
\usepackage{subfig}
\usepackage{float}
\usepackage{subfloat}

\title{DeFungi: Direct Mycological Examination of Microscopic Fungi Images}

\date{} 					% Or removing it

\author{{Camilo Javier Pineda Sopo}\\
	College of Engineering and Science\\
	Victoria University Sydney\\
	Sydney, NSW 2000, Australia\\
	\texttt{camilo.pinedasopo@students.vu.edu.au} \\
	\And
	Farshid Hajati\thanks{Corresponding Author, Address: 160 Sussex Street, Sydney, NSW 2000, Australia, Tel: +61 2 82653258} \\
	College of Engineering and Science\\
	Victoria University Sydney\\
	Sydney, NSW 2000, Australia\\
	\texttt{farshid.hajati@vu.edu.au} \\
	\AND
	Soheila Gheisari\\
	College of Engineering and Science\\
	Victoria University Sydney\\
	Sydney, NSW 2000, Australia\\
	\texttt{soheila.gheisari@vu.edu.au} \\
}

\renewcommand{\headeright}{}
\renewcommand{\undertitle}{}
\begin{document}
\maketitle

\begin{abstract}
Traditionally, diagnosis and treatment of fungal infections in humans depend heavily on face-to-face consultations or examinations made by specialized laboratory scientists known as mycologists. In many cases, such as the recent mucormycosis spread in the COVID-19 pandemic, an initial treatment can be safely suggested to the patient during the earliest stage of the mycological diagnostic process by performing a direct examination of biopsies or samples through a microscope. Computer-aided diagnosis systems using deep learning models have been trained and used for the late mycological diagnostic stages. However, there are no reference literature works made for the early stages. A mycological laboratory in Colombia donated the images used for the development of this research work. They were manually labelled into five classes and curated with a subject matter expert assistance. The images were later cropped and patched with automated code routines to produce the final dataset. This paper presents experimental results classifying five fungi types using two different deep learning approaches and three different convolutional neural network models, VGG16, Inception V3, and ResNet50. The first approach benchmarks the classification performance for the models trained from scratch, while the second approach benchmarks the classification performance using pre-trained models based on the ImageNet dataset. Using k-fold cross-validation testing on the 5-class dataset, the best performing model trained from scratch was Inception V3, reporting 73.2\% accuracy. Also, the best performing model using transfer learning was VGG16 reporting 85.04\%. The results presented by this paper are novel in the field of study. The statistics provided by the two approaches create an initial point of reference to encourage future research works to improve classification performance over the target classes and enhance the methodology used. Furthermore, the dataset built is published in Kaggle and GitHub to foster future research.\footnote{\url{https://www.kaggle.com/camilovu/lemm-pre-processed},  \url{https://www.kaggle.com/camilovu/lemm-raw}, \url{https://github.com/s4632865/lemm}} 
\end{abstract}

% keywords can be removed

\section{Introduction}
Fungi constituted as an individual kingdom of species from 1969 (Whitakker, 1969), have been leveraged and exploited in industrial practices such as pharmaceutical production, brewing, baking and others (Duddington, 1961). The acting of fungi as biological control agents is considered of paramount importance to ecology by contributing to the balance of earth’s ecosystems as recyclers and decomposers (Butt et al., 2001). The pathogenetic characteristics and diseases caused by some types of fungi have also been recognized by science as an essential matter to humanity. COVID-19 pandemic has produced relevant and novel statistical data about the coinfection of bacteria and fungi on critically ill patients. Zhou et al. (2020) outlined that 10\% of patients diagnosed with COVID-19 had a fungal and/or bacterial coinfection. The prevention of fungal infections is very relevant and estimate accounts for 1.5 to 2 million people dying from a fungal infection every year, primarily people acquiring fungal sepsis in hospitals and those with AIDS, according to Ketchell (2018). Mucormycosis infections, colloquially known as Black Fungi were discovered in 1815. Infection outbreaks by this fungus genus have been reported and controlled previously, as the causes have been identified in contaminated hospital medications or packaged foods, however in India and concurrently with the COVID-19 surge the infection rate has risen to 14 in every 100,000 people compared to 0.6 in every 100,000 people in Australia; 94\% of the mucormycosis related cases were associated with diabetic patients (John et al., 2021).

Diagnosis and classification of a fungal infection are made in a laboratory by a specialized biologist known as Mycologist, patient samples such as swabs, blood or scrabs of skin, hair, or nails are processed and cultured in controlled mediums for a range period of 28-31 days (Bosshard, 2011). During the evolutionary process of incubation and growth, the morphological characteristics of the fungi allow Mycologists to suggest a classification diagnosis to medical practitioners such as dermatologists to give early treatment to patients (Pihet et al., 2015). Early treated superficial fungal infections produce less painful and costly treatments to patients. Also, it lowers the percentage of mortality rates, as the 20\% increase is related to invasive Candidiasis for patients without early antifungal therapy (Kozel and Wickes (2014). The first phase of the mycological diagnosis process begins by conducting a direct examination (DE) of the sample through a microscope with augmentations ranging from 10x to 1000x with or without staining. Mycologists look for morphological patterns that can suggest unicellular yeasts, multi-cellular hyphae, or a combination of both; however, it is almost impossible to give a final diagnosis of the fungal infection through a DE due to the significant cell similarities between species (Robert, 2008). Therefore, Mycologists rely on cultured fungi growth to confirm the specific fungus genus and species. 

Mycologists are specialized medical laboratory scientists that live mainly in developed metropolitan areas where research and development funds nourish their work and income (Homei, 2006). Without specialized diagnosis coverage having an early or developed fungal infection, rural and urban areas will less likely and less timely get the proper treatment and medicine to counter the development of diseases. Computer Aided Diagnosis (CAD) has been accepted in modern medicine practices (Fiorini et al., 2019, Rafiei et al., 2021), the amount of research publications has been steadily increasing since 1969, constituted as the first practical CAD use by classifying pulmonary lesions, nowadays CAD regularly assist doctors in the interpretation and analysis of images such as x-rays and MRIs (Van Ginneken et al., 2011). Moreover, CAD systems improve medicine professional’s performance in diagnosing complex diseases such as breast cancer by increasing the number of biopsies for patients with malignant lesions and decreasing them for patients with benign lesions (Jiang et al., 1999). A future practical application of this research work would be an image classifier based in an Artificial Intelligence (AI) model implemented as a CAD system, this implementation can streamline and accelerate the initial diagnosis of a fungal infection by processing a DE image in a matter of minutes or seconds.

\subsection{Related Work}
Contemporary research works have processed microscopic images using neural networks to diagnose fungi species. (Zielinski et al., 2020) created a deep learning model to identify a range of Yeast fungi. The model is based on well-known convolutional neural networks (CNN) architecture and aids the accuracy by using the bag-of-words approach. They compared different classifier and convolutional combinations and obtained a 93\% accuracy model identifying nine different Candida types; the work relies on the microscopic examination of cultured samples, requiring the sample to have been processed and cultured by a specialized biologist. Also, their system was trained with a public reservoir of data around Candida yeasts that made the training, validation, and testing classified as output nine fungi species. (Hao et al., 2019) created a deep neural network aiding gynaecological tests identify Candida Albicans, a single species yeast type fungus based on interesting practices including image preprocessing using maximum inter-class variation segmentation and pre-classifier processing using morphological methods such as concave contour detection. The study enhances traditional CNN and morphological methods of identifying fungi and achieved an accuracy of 93\%. Similarly, Cuervo et al. (2019) created a system based on CNN using MATLAB as the software tool. Their model was able to classify four species of Fusarium mould fungi based on a final culture of the infection sample. Their work relies on very high-definition input samples but carries an interesting image preprocessing to perform feature extraction. On the other hand, their work achieved a modest accuracy of 69.51\% due to the short amount of test, training, and validation data alongside acknowledged limitations from the input entities. 

Zieliński et al. (2020) developed a system using deep neural networks and Support Vector Machine (SVM) to classify nine fungi species that produce candidacies infections that are considered standard worldwide. Their study team acquired fungi strains from an American laboratory and then cultured them to create 180 high-resolution images using professional-grade equipment manually. The study’s outcome was to replace the last stage of traditional mycological diagnosis known as biochemical tests, saving up to 3 days. Tahir et al. (2018) study included the donation of 40,800 images created by an automated mechanism attached to a microscope. Images were sourced from cultured samples in a laboratory. They accurately classified the dataset based on external factors such as light sources and camera focus measure methods. Their fungal category includes yeast-like and mould-like types of fungi. Even spores cultured from the soil and the air using an air sampling unit were classified. The authors reported a system’s accuracy of 94.8\% identifying five species and claim a better performance when comparing Hao et al. (2019) 93.26\% accuracy identifying a single class. 

Complementary research to the traditional deep learning frameworks has been done on different fronts; transfer learning (TL) tries to circumvent the performance degradation of conventional deep models with limited datasets by re-using pre-trained networks. Research done by Mital et al. (2020) performed benchmarking comparisons over a set of well-known CNN models that were pre-trained on a relevant set to the target problem: classify Fusarium fungi species. Their results exhibited state-of-the-art performance statistics and crowned the Inception V2 model as the best balance between accuracy and computational cost.
No other studies have worked on superficial fungal infection classification using DE images to the best of our knowledge. This research provides novel results in the classification of superficial fungal infections using deep models. The deep models to be benchmarked have a set of tailored automated and manual image pre-processing modules to standardize, enhance and augment the input dataset where applicable. The experiments to be conducted will be done over a 5-class dataset.

The significance of this research covers various folds;

\begin{itemize}
\item Firstly, to date, no literary works are exploring the classification performance of machine learning models based on images from fungal infections retrieved from DE procedures. The gap is thought to be tied to the fact that the target class image dataset is hard to collect and process; the involved steps of labelling, curating, and processing require SME assistance.

\item Secondly, the benchmark results presented by this research work will establish an initial reference point for future works. The modelling and processing presented can be improved using the same benchmarked models, or the performance of a different model or set of models can be gauged.
\end{itemize}

\section{DeFungi Dataset}
\label{sec:headings}

This study aims to fill the gap encountered in fungi classification from DE microscopic images. For this purpose, we have used a dataset donated by LEMM \footnote{\url{https://www.leticiasopomicologia.com}}, including around 3,000 uncurated and unlabeled images from superficial fungal infections caused by yeasts, moulds, or dermatophyte fungi. The following bullet points outline relevant information and statistics of the input data collected to conduct the experiments.

\begin{itemize}
\item All data are DE microscopic images sourced from Sony’s DSC W830 compact camera.
\item The type of input data is .jpg files.
\item The size of the input data varies from 120KB up to 8,000+KB.
\item The resolution of the input data varies from 640×480 pixels up to 5152×3864+ pixels.
\item The total amount of non-curated, non-labelled data is 3,025 images.
\item The total amount of labelled data is 660 images.
\item No pre-processing has been made by LEMM on any of the input images.
\item No personal information is correlated or included in any of the input images.
\item Raw input data was being continually added into the dataset until February 2021, after that, the data was labelled, curated, and preprocessed to favour the readiness for experimentation.
\end{itemize}

\begin{figure*}[htb]
	\centering
		\includegraphics[scale=1]{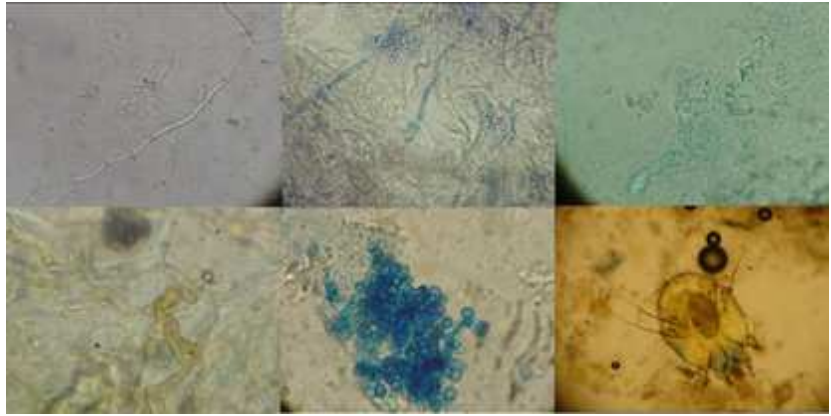}
	\caption{Collage of raw images in the dataset}
	\label{fig1}
\end{figure*}

Figure 1 shows a collage of the different fungi, yeasts, and other microorganisms present in the raw set of images. To train, tune, and deliver a state-of-the-art performance-based in a deep learning model, the dataset size is of paramount importance. Therefore, five different fungi types were selected as follows:

\begin{itemize}
\item Tortuous septate hyaline hyphae (TSH) compatible with dermatophytes shown in Figure 2(a).
\item Beaded arthroconidial septate hyaline hyphae (BASH) compatible with dermatophytes shown in Figure 2(b). 
\item Groups or mosaics of arthroconidia (GMA) compatible with dermatophytes shown in Figure 2(c). 
\item Septate hyaline hyphae with chlamydioconidia (SHC) compatible with moulds shown in Figure 2(d). 
\item Broad brown hyphae (BBH) compatible with dematiaceous moulds shown in Figure 2(d). 
\end{itemize}

Out of the 3,000 raw images and with SME guidance, the following distribution was obtained: TSH (227 images), BASH (117 images), GMA (36 images), SHC (144 images), BBH (75 images).

\begin{figure*}[htb]
	\centering
		\includegraphics[scale=1]{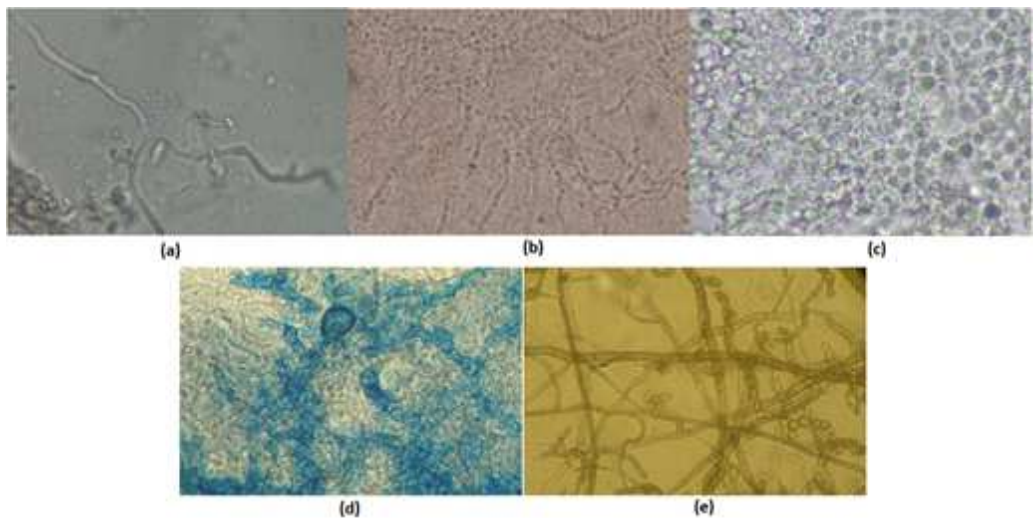}
	\caption{Five samples of fungi types in the DeFungi dataset. (a) Tortuous septate hyaline hyphae (TSH), (b) Beaded arthroconidial septate hyaline hyphae (BASH), (c) Groups or mosaics of arthroconidia (GMA), (d) Septate hyaline hyphae with chlamydioconidia (SHC), (e) Broad brown hyphae (BBH).}
	\label{fig2}
\end{figure*}

\subsection{Data Preprocessing}
The raw images cannot be fed directly into the deep learning models to be benchmarked. Additional filtering, patching and preprocessing procedures were executed over the raw dataset to produce an academic quality dataset. One of the challenges addressed by this research work was having a dataset of raw images. Besides being of different sizes and quality, the raw images possess a relevant amount of noise and alien artefacts that will need to be filtered out. Therefore, a set of manual and automated tasks were carried to curate and filter the labelled dataset provided, as explained in the following paragraphs.

All images need to be patched, irrespective of the input image resolution. The images were cut into squared patches by a proprietary python code explicitly designed for this task. The patch size can be chosen at runtime. However, the selected value to perform the experiments was 500×500 pixels. The image shown in the left part of Figure 3 has a 5152×3864 pixel resolution. By automatically patching the raw image into 500×500 patches, a set of 88 patches are produced. The ones squared in red from Figure 3 are six randomly selected patches; those on top are relevant patches containing fungal cells, the ones at the bottom are non-relevant patches that will have to be ruled out.

\begin{figure*}[htb]
	\centering
		\includegraphics[scale=1]{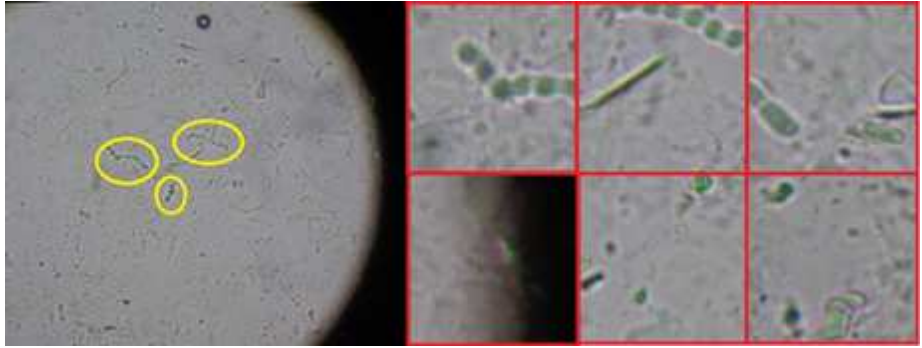}
	\caption{A sample raw image with extracted patches.}
	\label{fig3}
\end{figure*}

The patches need to be filtered; the relevant content of each image can be defined as a set of multi-cellular morphological patterns. Patches including alien artefacts such as black microscope lens contour, blank/colored spaces with no cells and alien artefacts such as cream and sunscreen should be ruled out. The black contours were filtered out automatically by a proprietary algorithm explicitly designed for this task. As an example, the raw image shown in Figure 4 will produce three rows of patches from the highlighted area on top and one patch from the highlighted area on the bottom right. Those patches will be ruled out automatically by the designed algorithm based on contrast radio analysis.

\begin{figure*}[htb]
	\centering
		\includegraphics[scale=1]{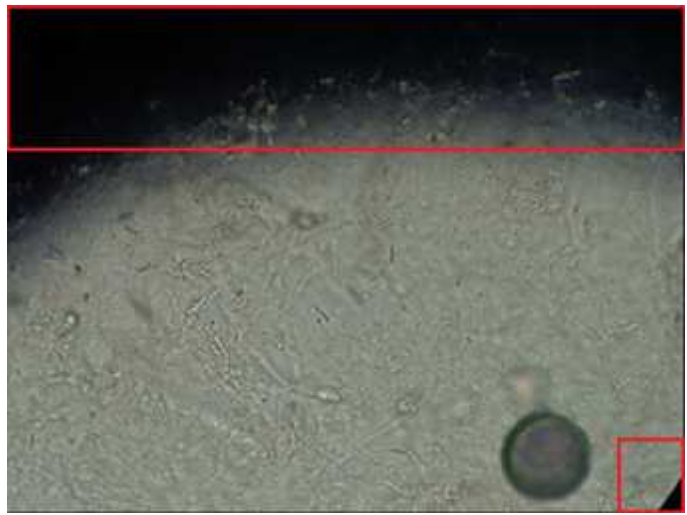}
	\caption{A TSH image with notorious black lens contour.}
	\label{fig4}
\end{figure*}

The blank and/or colored spaces were filtered out automatically by a proprietary algorithm, a manual quality check was needed over the code testing to assure patches with relevant data were not being discarded. For example, the image shown in Figure 5 has a large set of blank areas. After black contour preprocessing is done, the blank areas are discarded with a much lower contrast ratio to avoid discarding areas with fungal cells.

\begin{figure*}[htb]
	\centering
		\includegraphics[scale=1]{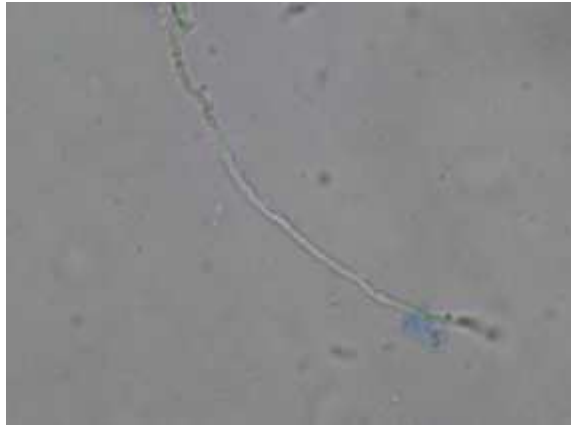}
	\caption{A TSH image with large blank areas.}
	\label{fig5}
\end{figure*}

Final filtering needed to be done by manually selecting patches with relevant fungal cells vs alien artefacts. This process was done with assistance from LEMM for a set of 50 patches; the filtering was finalized by manually ruling out non-relevant patches. The image shown in Figure 6 has the fungal cells highlighted in yellow; the patches highlighted in red were filtered automatically by the preprocessing algorithms described above. However, the ones highlighted in blue are alien artefacts that could be interpreted as fungal cells. Therefore, the rule out should be done manually or by a more advanced algorithm. The patches highlighted in lime-green are fungal cells that correspond to the zone highlighted in yellow.

\begin{figure*}[htb]
	\centering
		\includegraphics[scale=1]{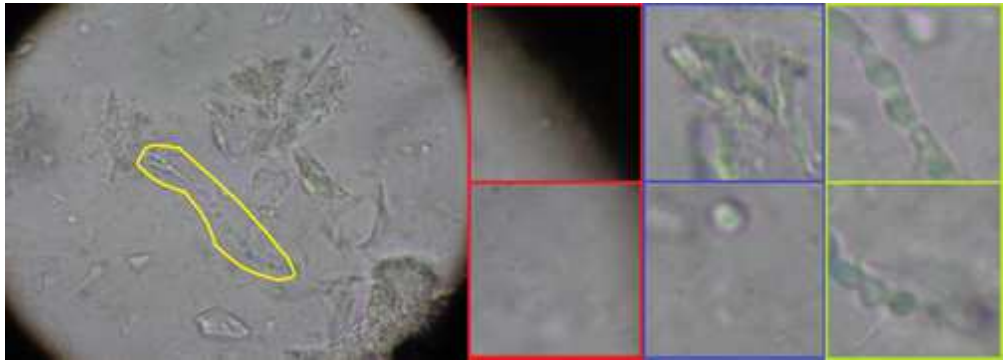}
	\caption{A BASH image with noise and blank areas.}
	\label{fig6}
\end{figure*}

The final amount of preprocessed data ready to be used for experiments is 9,114 images. The data was split following standardized and widely used ratios, 85/15 when Train/Validation is done and 76.5/13.5/10 when Train/Validation/Testing is done.

\section{Classification}
\label{sec:headings}
To classify the build dataset images, we have applied ResNet50, Inception V3, and VGG16 models. These models are explained in the below subsections.

\subsection{ResNet50}
The Resnet family has existed for almost 6 years now when the first models were introduced in 2015. Ever since different versions of the initial model have been published and due to the enhanced capabilities, it possesses it has been widely used to perform classification experiments on a heterogeneous set of target classes. Common DL NN problems such as overfitting and exploding gradient have been circumvented in literature works by stacking layers to the models, which ultimately results in better accuracy performance at a cost of other degradations. ResNet50 introduces residual mapping fit between three layers, this is done by creating shortcuts between a set of layers by using an identity function.

\subsection{Inception V3}
Inception version 3 model (Purva, 2020), also known as GoogLeNet was created in 2014 contemporary to VGG-16. Unlike VGG16 this model is less computational complex, yet it has reported better performance results in different image classification scenarios. The better performance is believed to be attributed to the inception module that performs feature extraction tasks with different size filters.

\subsection{VGG16}
VGG16 was introduced in 2014 (Purva, 2020). This model still provides best performing results for image recognition studies, its computational complexity is recognized among the largest with around 138 billion parameters.

\section{Results}
\label{sec:headings}

The accuracy and loss function performance experiments were executed for all models during the train and validation phases. These were executed in two approaches, namely, using and not using transfer learning (TL). The results are presented as plotted graphs over a lapse of epochs. The experiments were run over the constructed 5-class dataset. All experiments were executed with the following configurations:

\begin{itemize}
\item Total dataset size (5 classes) = 2,500
\item Train-Validation dataset ratio: 85/15
\item Train batch size: 26
\item Validation batch size: 70
\item Using dataset augmentation in the train set
\item Optimizer: Adam
\item Learning rate: 0.00001
\item Loss function: Categorical cross-entropy
\item Training steps per epoch: 80
\item Validation steps per epoch: 5
\item Epochs using transfer learning: 100
\item Epochs not using transfer learning: 200
\end{itemize}

The results of the experiments are shown in Figures 7-9.

\begin{figure*}[htb]
   \centering
    \subfloat[]{\includegraphics{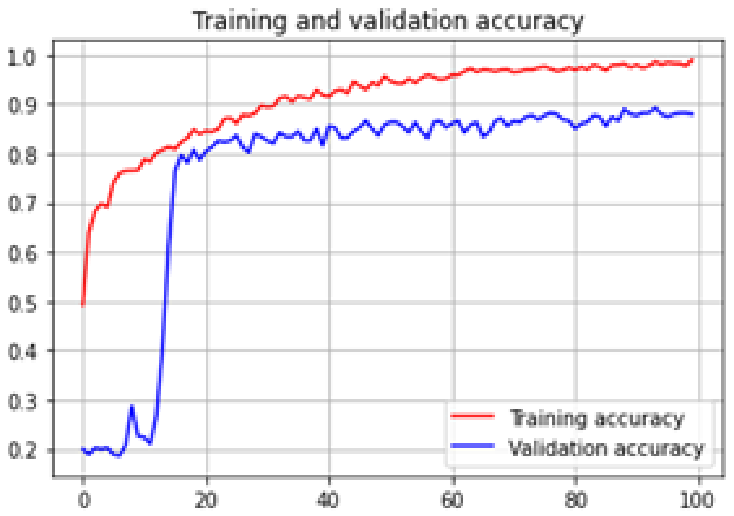}}
    \subfloat[]{\includegraphics{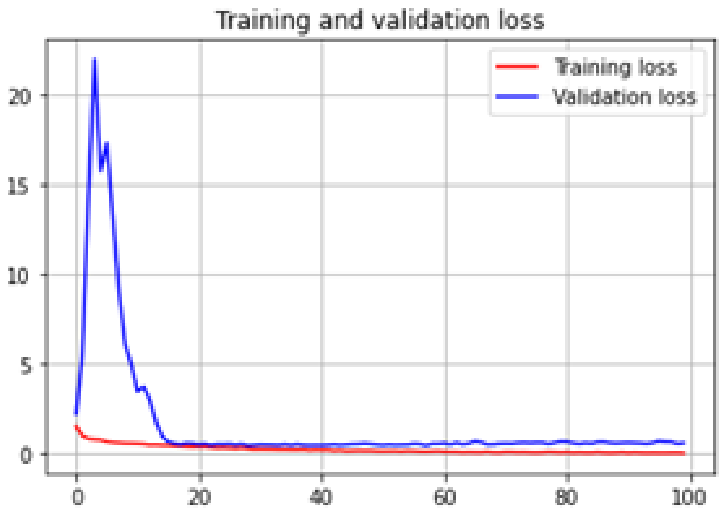}}
    \caption{ResNet50 with transfer learning train-validation accuracy.}%
    \label{fig7}
\end{figure*}

\begin{figure*}[htb]
   \centering
    \subfloat[]{\includegraphics{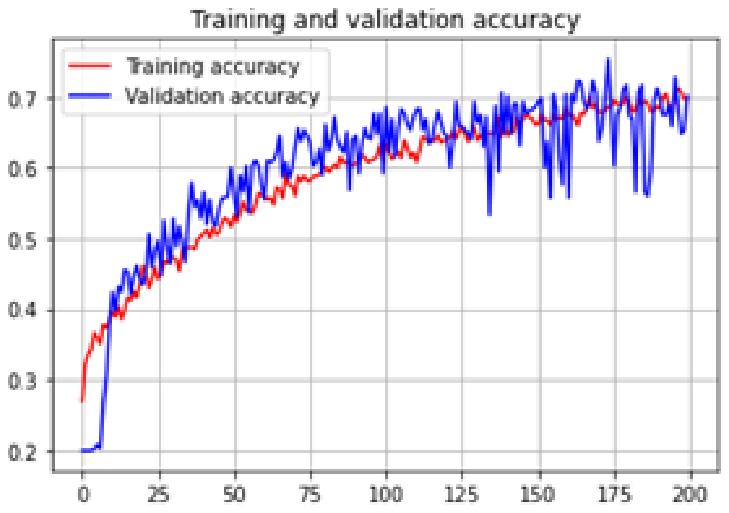} }%
    \subfloat[]{\includegraphics{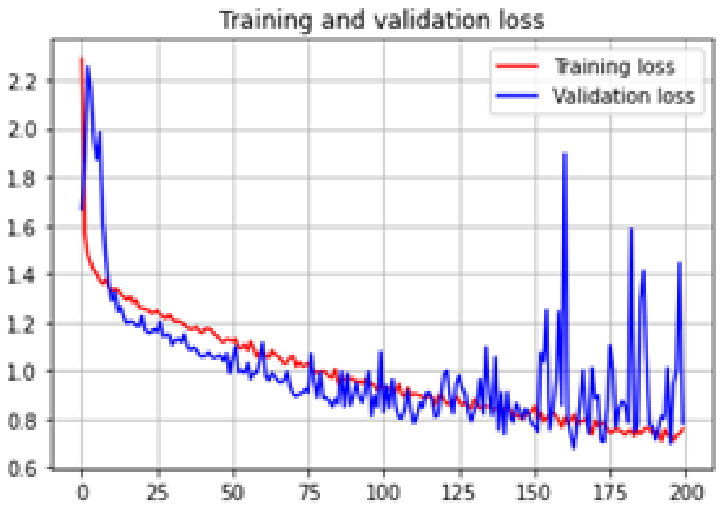}}%
    \caption{ResNet50 without transfer learning train-validation accuracy.}%
    \label{fig8}%
\end{figure*}

\begin{figure*}[htb]
   \centering
    \subfloat[]{{\includegraphics[scale=1]{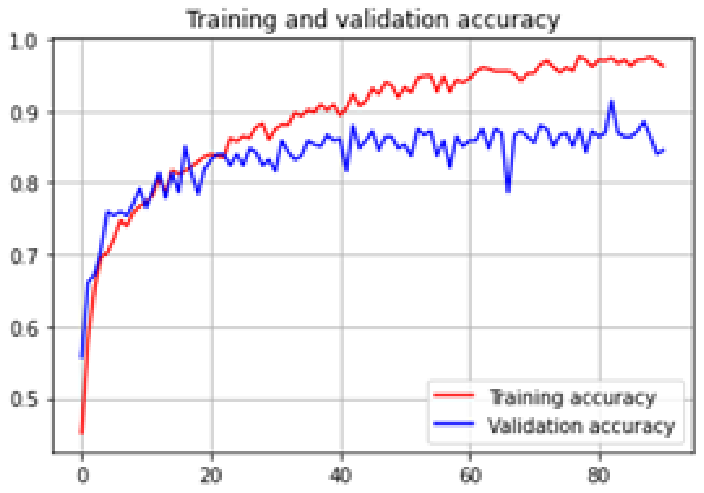} }}%
    \subfloat[]{{\includegraphics[scale=1]{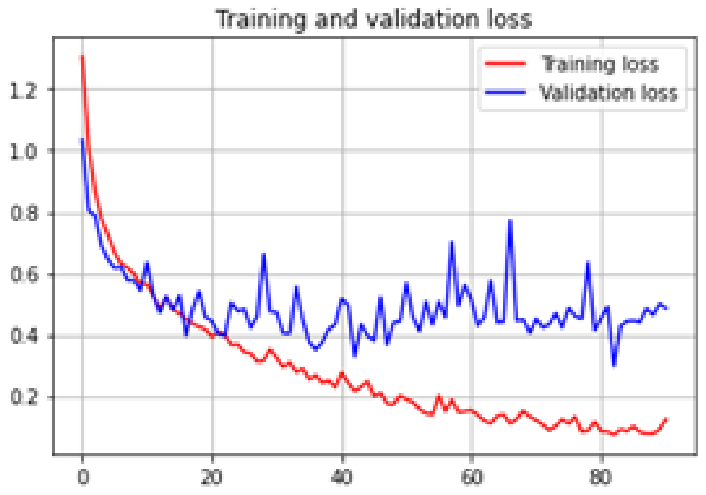}}}%
    \caption{VGG16 with transfer learning train-validation accuracy.}%
    \label{fig9}%
\end{figure*}

The training and validation error and loss performance results are “easy” estimations of prediction accuracy for the trained models; however, cross-validation testing techniques have been established as a reference for presenting classifier performance literature contributions. As stated by Anguita et al. (2012) and Fushiki (2011), k-fold cross-validation is the favoured method to gauge the prediction/classification performance of ML models, firstly because the bias and trends of the error and cross-validation in folds are correlated and because it is one of the most effective and computational friendly methods to deliver the performance of test results.
The number of folds to use is a matter of scientific study related to the dataset and its properties, however, Fushiki (2011) have studied and benchmarked the trade-off using different fold values for a set of renowned datasets. For this research work, the number of folds to be used as suggested by the thesis supervisor will be 10. The validation experiments will be done only for the models using transfer learning.
All testing experiments were executed with the following configurations:

\begin{itemize}
\item K-folds = 10
\item Total dataset size (5 classes) = 2.500
\item Train/Validation/Test dataset ratio: 76.5/13.5/10
\item Dataset shuffling being done by: K-fold
\item Train batch size: 24
\item Validation batch size: 56
\item Test batch size: 45
\item Using dataset augmentation in the train set
\item Optimizer: Adam
\item Learning rate: 0.00001
\item Loss function: Categorical cross-entropy
\item Steps per epoch: 80
\item Validation steps: 6
\item Epochs using transfer learning: 100
\item Epochs not using transfer learning: 200
\item Callback function stops the training after 8 consecutive steps without loss function optimization.
\end{itemize}

The results of the experiments are shown in Figures 10-12.

\begin{figure*}[htb]
   \centering
    \subfloat[]{{\includegraphics[scale=1]{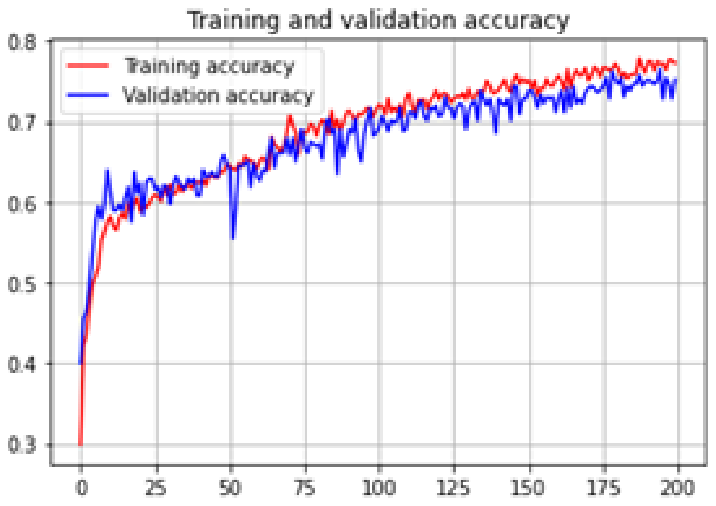}}}%
    \subfloat[]{{\includegraphics[scale=1]{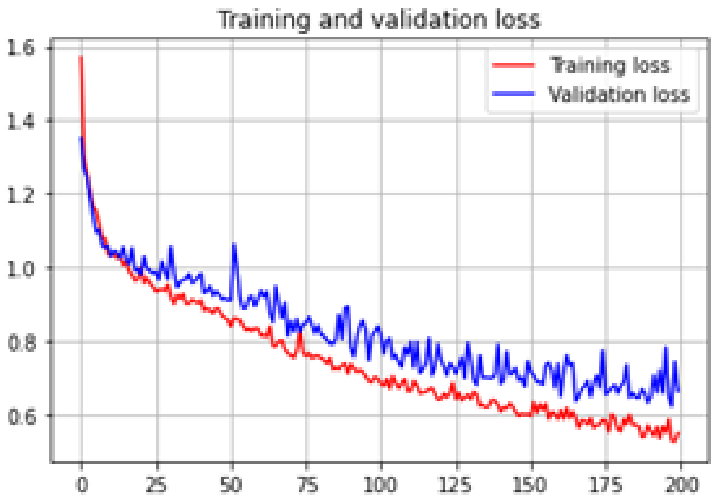}}}%
    \caption{VGG16 without transfer learning train-validation accuracy.}%
    \label{fig10}%
\end{figure*}

\begin{figure*}[htb]
   \centering
    \subfloat[]{{\includegraphics[scale=1]{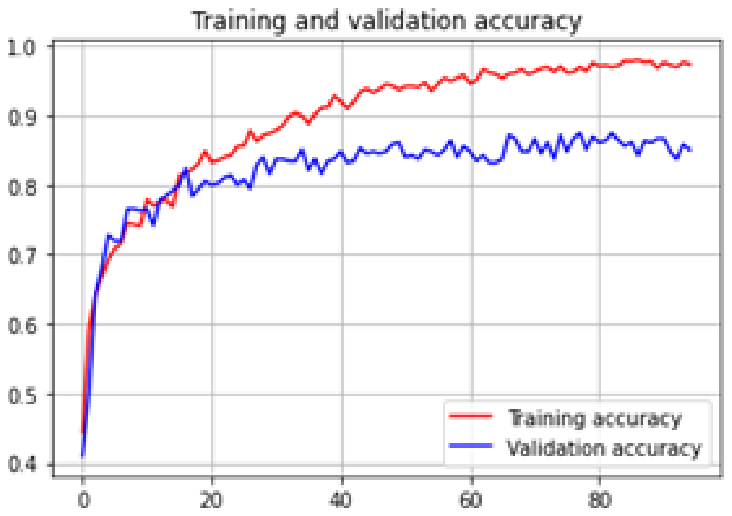} }}%
    \subfloat[]{{\includegraphics[scale=1]{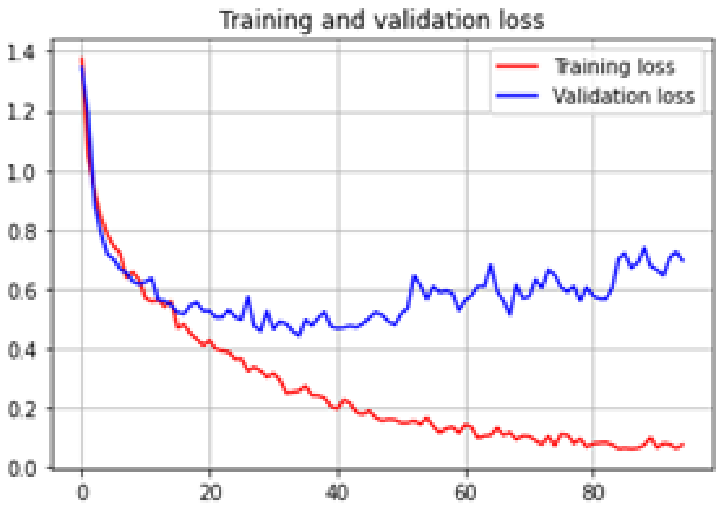}}}%
    \caption{Inception V3 with transfer learning train-validation accuracy.}%
    \label{fig11}%
\end{figure*}

\begin{figure*}[htb]
   \centering
    \subfloat[]{{\includegraphics[scale=1]{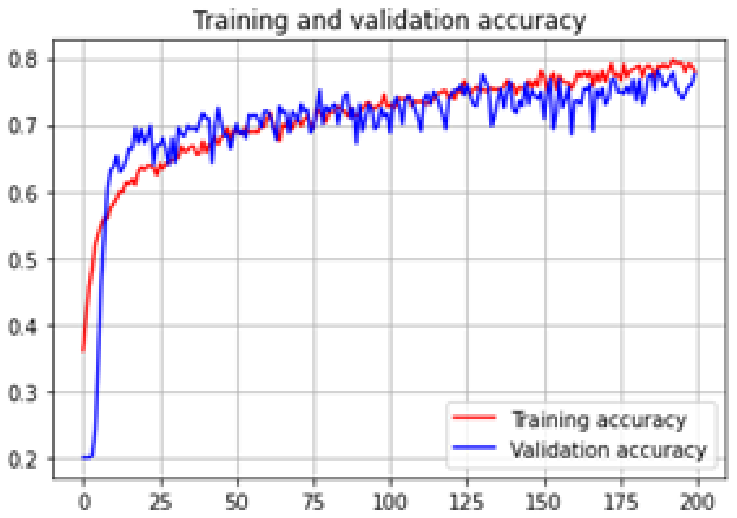} }}%
    \subfloat[]{{\includegraphics[scale=1]{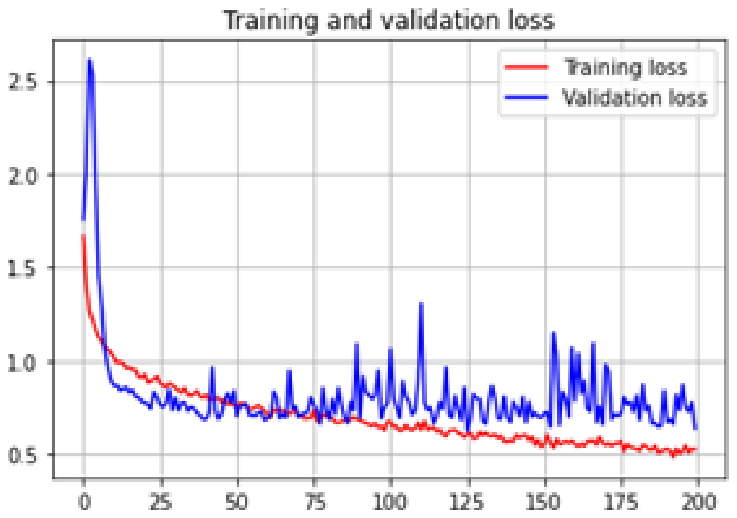}}}%
    \caption{Inception V3 without transfer learning train-validation accuracy.}%
    \label{fig12}%
\end{figure*}

All models reach accuracies greater or equal than 80\% before epoch 20 (green vertical line), thereafter, the most optimal trend is exhibited by the ResNet50 model achieving accuracies close to 90\% by epoch 90, VGG16 and Inception V3 oscillate and show a steady trend after epoch 40. For reference, the horizontal orange line corresponds to the highest 10-fold test accuracy reported by VGG16 using transfer learning, which is 85.04\%. VGG16 reported accuracies greater or equal to that value around epoch 30, whereas the other two models reported them around epoch 40. Additionally, VGG16 reported the largest number of accuracy points over the orange line compared to its counterparts.
The results documented in Table 1 must be coherent in nature to those reported by the train-validation experiments, in fact, they were, the overall best performing model is VGG16, it held the highest test accuracy performance with the lowest values in standard deviation and loss. As a concluding remark, this research work acknowledges the fact that the highest accuracies with less computational cost could have been achieved by performing model tuning and more advanced dataset preprocessing.

\begin{table}[!htbp]
\centering
\caption{10-fold test accuracy performance comparison of the models using transfer learning.}
\begin{tabular}{*9c}
\toprule
\multicolumn{3}{c}{ResNet50} &  \multicolumn{3}{c}{VGG16} & \multicolumn{3}{c}{Inception V3}\\
\midrule
Fold&Loss&Accuracy &Fold&Loss&Accuracy&Fold&Loss&Accuracy\\
1   &  0.894 & 85.199\%& 1   &  0.604 & 84.799\%&1   &  0.815 & 79.600\% \\
2   & 0.968  &  83.600\%  & 2   & 0.674  &  82.400\%  &2   & 0.399  &  87.999\% \\
3   & 0.865   &  85.600\%  & 3   & 0.447   &  88.800\%  &3   & 0.757   &  80.000\%  \\
4   & 0.715  &  80.000\%  & 4   & 0.523  &  85.600\%  &4   & 0.762  &  83.600\%  \\
5   & 0.814  &  82.800\%  & 5   & 0.68  &  83.999\%  &5   & 0.742  &  76.399\%  \\
6   & 1.118   & 83.999\%    & 6   & 0.573   & 83.200\% &6   & 0.64   & 82.800\%   \\
7   & 1.099  & 80.000\%   & 7   & 0.498  & 86.799\%   &7   & 0.719  & 84.799\%   \\
8   & 0.68  & 83.999\%   & 8   & 0.76  & 84.399\%   &8   & 0.946  & 81.999\%  \\
9   &  0.692  &  84.399\%   & 9   &  0.748  &  83.600\%   &9   &  0.763  &  82.400\%   \\
10   & 0.927  &  80.800\%  & 10   & 0.44  &  86.799\%  &10   & 0.406  &  88.400\%  \\ 
\midrule
\multicolumn{2}{c}{Average Loss:}& 0.877&\multicolumn{2}{c}{Average Loss:}&0.5947& \multicolumn{2}{c}{Average Loss:}&0.6949\\
\midrule
\multicolumn{2}{c}{Average Accuracy:}& 85.040\%& \multicolumn{2}{c}{Average Accuracy:}&83.040\%& \multicolumn{2}{c}{Average Accuracy:}&82.800\%\\
\bottomrule
\multicolumn{2}{c}{Standard Deviation:}& 1.969\%& \multicolumn{2}{c}{Standard Deviation:}&1.861\%& \multicolumn{2}{c}{Standard Deviation:}&3.505\%\\
\bottomrule
\end{tabular}
\end{table}

The results obtained by developing experiments with models trained from scratch confirm and validate the great utility of transferring knowledge in terms of weight and bias initialization to enhance performance. When not using transfer learning, Inception V3 being the best performing model reported 73.28\% over 200 epochs compared to 85.04\% over 100 epochs reported by VGG16, similarly the average loss function value reported by the former was 0.710 compared to 0.594 by the latter. Both improvements were achieved by using transfer learning in shorter elapsed time periods, hence lower computational complexity requirements. Future research works could further improve the reference values presented by this study by performing a specialized model tuning or a more refined image pre-processing on the published dataset.
Inception V3 outperforms all other models as early as the 20th epoch, additionally, the validation accuracy values surpass the 73.28\% value much earlier than all others while showing similar oscillation amplitudes to VGG16. All models present a tendency to optimize accuracy performance over time, though the oscillations exhibited by ResNet50 blur the model’s performance completely.
Comparison details of the test performance not using transfer learning are in Table 2. The last layer for models trained from scratch used an additional dropout step after the 512-dense layer, the adding of the step and general configuration of the top layer has a significant impact on the model’s performance. As a concluding remark from the discussion, this research work acknowledges the fact that the highest accuracies with less computational cost could have been achieved by performing model tuning and more advanced dataset preprocessing.

\begin{table}[!htbp]
\centering
\caption{10-fold test accuracy performance comparison of the models without transfer learning.}
\begin{tabular}{*9c}
\toprule
\multicolumn{3}{c}{ResNet50} &  \multicolumn{3}{c}{VGG16} & \multicolumn{3}{c}{Inception V3}\\
\midrule
Fold&Loss&Accuracy &Fold&Loss&Accuracy&Fold&Loss&Accuracy\\
1   &  0.905 & 62.40\%& 1   &  0.673 & 72.40\%&1   &  0.738 & 71.60\% \\
2   & 0.97  &  62.00\%  & 2   & 0.764  &  74.40\%  &2   & 0.808  &  70.00\% \\
3   & 0.77   &  68.80\%  & 3   & 0.716   &  68.80\%  &3   & 0.633   &  74.00\%  \\
4   & 0.741  &  66.80\%  & 4   & 0.709  &  68.80\%  &4   & 0.616  &  74.00\%  \\
5   & 0.796  &  70.00\%  & 5   & 0.776  &  68.00\%  &5   & 0.722  &  75.20\%  \\
6   & 0.989   & 62.40\%    & 6   & 0.675   & 73.20\% &6   & 0.747   & 72.80\%   \\
7   & 0.786  & 68.80\%   & 7   & 0.627  & 76.80\%   &7   & 0.705  & 74.80\%   \\
8   & 0.77  & 70.80\%   & 8   & 0.646  & 71.60\%   &8   & 0.699  & 73.60\%  \\
9   &  0.927  &  61.60\%   & 9   &  0.724  &  69.60\%   &9   &  0.86  &  71.20\%   \\
10   & 0.854  &  67.60\%  & 10   & 0.745  &  71.20\%  &10   & 0.58  &  75.60\%  \\ 
\midrule
\multicolumn{2}{c}{Average Loss:}& 0.8508 & \multicolumn{2}{c}{Average Loss:}&0.7055& \multicolumn{2}{c}{Average Accuracy:}&0.7108\\
\midrule
\multicolumn{2}{c}{Average Accuracy:}& 66.120\%& \multicolumn{2}{c}{Average Accuracy:}&71.480\%& \multicolumn{2}{c}{Average Accuracy:}&73.280\%\\
\bottomrule
\multicolumn{2}{c}{Standard Deviation:}& 3.450\%& \multicolumn{2}{c}{Standard Deviation:}&2.660\%& \multicolumn{2}{c}{Standard Deviation:}&1.751\%\\
\bottomrule
\end{tabular}
\end{table}

\section{Conclusions}
\label{sec:headings}
The research work done by this study has demonstrated that state-of-the-art accuracy performance can be achieved by classifying novel target class images by using DL CNN models. The results reported by the two approaches used to outline the advantageous use of transfer learning, the accuracy could be boosted by 20\% and the computational complexity in terms of execution time could be decreased significantly. The contribution to the field of study is two-fold, firstly, setting an initial reference benchmark comparison between recent high-performing and well know DL CNN models and secondly, by publishing into an open-source repository the raw and pre-processed dataset used to encourage future research. Research initiatives are usually challenged by the lack of data, even in raw form, therefore, many of these end up being withdrawn. DE images from superficial fungal infections in humans were no exception to the common challenge. In conclusion, any research work wanting to provide statistical information over a blank space class will need a sponsor and an appropriate project plan with sufficient time frames to build an academic quality grade dataset. The performance results obtained for all models, over the two approaches used and having done all experiments for each approach using the same runtime configurations rendered best performers. Inception V3 demonstrated the best accuracy performance and the best trend to optimal convergence than its counterparts when not using transfer learning, the architectural nature of this model maximizes the flow of information, this incurs in increased depth and width, hence feature maps as well, not surprisingly Inception V3 reported the longest execution elapsed times compared with other models when infrastructure allocation was guaranteed; 

\section*{Acknowledgement}
The achievement of the results could not have been done without LEMM's sponsorship and counsellorship. LEMM dedicated countless hours supporting the development of this research work helping label, curating, and advising over image preprocessing techniques to be implemented, additionally, they helped to perform quality checks on classification performance and dataset readiness.

\section*{References}
%\bibliographystyle{unsrtnat}
%\bibliography{references} 
Anguita, D., Ghelardoni, L., Ghio, A., Oneto, L., and Ridella, S. (2012, April). The 'K' in K-fold Cross Validation. In ESANN (pp. 441-446).

Bonifaz Trujillo J. Mucormicosis y entomoftoromicosis (zigomicosis). Micologia Medica Basica, 5e. McGraw-Hill. 2015.

Bonifaz A, Tirado-Sanchez A, Hernandez-Medel ML, Araiza J, Kassack JJ, Del Angel-Arenas T, Moises-Hernández JF, Paredes-Farrera F, Gomez-Apo E, Trevino-Rangel RJ, Gonzalez GM, 2020. Mucormycosis at a tertiary-care center in Mexico. A 35-year retrospective study of 214 cases. Mycoses. 2021 Apr;64(4):372-380.

Bosshard, P.P., 2011. Incubation of fungal cultures: how long is long enough? Mycoses, 54(5), pp.e539-e545.

Butt, T.M., Jackson, C. and Magan, N. eds., 2001. Fungi as biocontrol agents: progress problems and potential. CABI.

Canziani, A., Paszke, A. and Culurciello, E., 2016. An analysis of deep neural network models for practical applications. arXiv preprint arXiv:1605.07678.

Cuervo, S., Bolanos, F., Vallejo, M. and Mesa-Arango, A.C., 2019, October. Fusarium species identification by means of digital signal processing. In 2019 IEEE 4th Colombian Conference on Automatic Control (CCAC) (pp. 1-5).

Duddington, C.L., 1961. Microorganisms as allies. The industrial use of fungi and bacteria. Microorganisms as allies. The industrial use of fungi and bacteria.

Fiorini, S., Hajati, F., Barla, A. and Girosi, F., 2019. Predicting diabetes second-line therapy initiation in the Australian population via time span-guided neural attention network. PloS one, 14(10), p.e0211844.

Fushiki, Tadayoshi. “Estimation of prediction error by using K-fold cross-validation.” Statistics and Computing 21.2 (2011): 137-146.

Hao, R., Wang, X., Zhang, J., Liu, J., Du, X. and Liu, L., 2019, March. Automatic Detection of Fungi in Microscopic Leucorrhea Images Based on Convolutional Neural Network and Morphological Method. In 2019 IEEE 3rd Information Technology, Networking, Electronic and Automation Control Conference (ITNEC) (pp. 2491-2494).

He, K., Zhang, X., Ren, S., and Sun, J. (2016). Deep residual learning for image recognition. In Proceedings of the IEEE conference on computer vision and pattern recognition (pp. 770-778).

Homei, A., 2006. Medical mycology development and epidemiology in the USA, UK and Japan. Medical Mycology, 44(Supplement 1), pp.S39-S54.

Jiang, Y., Nishikawa, R.M., Schmidt, R.A., Metz, C.E., Giger, M.L. and Doi, K., 1999. Improving breast cancer diagnosis with computer-aided diagnosis. Academic radiology, 6(1), pp.22-33.

John, T. M., Jacob, C. N., and Kontoyiannis, D. P. (2021). When Uncontrolled Diabetes Mellitus and Severe COVID-19 Converge: The Perfect Storm for Mucormycosis. Journal of fungi (Basel, Switzerland), 7(4), 298.

Kozel, T. R., and Wickes, B. 2014. Fungal diagnostics. Cold Spring Harbor perspectives in medicine, 4(4).

McLean, A.C., Valenzuela, N., Fai, S. and Bennett, S.A., 2012. Performing vaginal lavage, crystal violet staining, and vaginal cytological evaluation for mouse estrous cycle staging identification. JoVE (Journal of Visualized Experiments), (67), p.e4389.

Mital, M.E., Tobias, R.R., Villaruel, H., Maningo, J.M., Billones, R.K., Vicerra, R.R., Bandala, A. and Dadios, E., 2020, November. Transfer Learning Approach for the Classification of Conidial Fungi (Genus Aspergillus) Thru Pre-trained Deep Learning Models. In 2020 IEEE REGION 10 CONFERENCE (TENCON) (pp. 1069-1074).

Pihet, M., Clement, N., Kauffmann-Lacroix, C., Nail-Billaud, S., Marot, A., Pilon, F. and Robert, R., 2015. Diagnosis of dermatophytosis: an evaluation of direct examination using MycetColor® and MycetFluo. Diagnostic Microbiology and Infectious Disease, 83(2), pp.170-174.

Purva Huilgol, Analytics Vidhya, 2020, digital image, Layout of the VGG-16 model, accessed 18 January 2021,
<https://www.analyticsvidhya.com/blog/2020/08/top-4-pre-trained-models-for-image-classification-with-python-code/>

Purva Huilgol, Analytics Vidhya, 2020, digital image, Inception V2 model, accessed 18 January 2021, <https://www.analyticsvidhya.com/blog/2020/08/top-4-pre-trained-models-for-image-classification-with-python-code/>

Purva Huilgol, Analytics Vidhya, 2020, digital image, EfficientNetB0 model, accessed 18 January 2021, <https://www.analyticsvidhya.com/blog/2020/08/top-4-pre-trained-models-for-image-classification-with-python-code/>

Rafiei, A., Rezaee, A., Hajati, F., Gheisari, S. and Golzan, M., 2021. SSP: Early prediction of sepsis using fully connected LSTM-CNN model. Computers in biology and medicine, 128, p.104110.

Robert, R. and Pihet, M., 2008. Conventional methods for the diagnosis of dermatophytosis. Mycopathologia, 166(5-6), pp.295-306.

Sopo Prada, L. (2020). Interviewed by Camilo Pineda via Google Meet, January 15.

Sopo Prada, L. (2020). LEMM procedure manual 2019-2020. Diagram elaborated by Manuela Pineda.

Tahir, M.W., Zaidi, N.A., Rao, A.A., Blank, R., Vellekoop, M.J. and Lang, W., 2018. A fungus spores dataset and a convolutional neural network based approach for fungus detection. IEEE transactions on nanobioscience, 17(3), pp.281-290.

TensorFlow Core v2.4.1, 2020, Module: tf.keras.applications, accessed 23 April 2021, <https://www.tensorflow.org/apidocs/python/tf/keras/applications>

The TensorFlow Authors, ML Zero to Hero course, Course 2, Part 8, accessed 09/02/2021, <http://bit.ly/2lXXdw5>

Van Ginneken, B., Schaefer-Prokop, C.M. and Prokop, M., 2011. Computer-aided diagnosis: how to move from the laboratory to the clinic. Radiology, 261(3), pp.719-732.

Zielinski, B., Sroka-Oleksiak, A., Rymarczyk, D., Piekarczyk, A., Brzychczy-Wloch, M., 2020. Deep learning approach to describe and classify fungi microscopic images.

Zhou, P., Liu, Z., Chen, Y., Xiao, Y., Huang, X. and Fan, X.G., 2020. Bacterial and fungal infections in COVID-19 patients: A matter of concern. Infection Control and Hospital Epidemiology, pp.1-2.

\end{document}